\documentclass{elsart}
\usepackage{epsfig}
\usepackage{psfrag}
\usepackage{amssymb}
\usepackage{amsmath}
\usepackage{bm}
\usepackage{graphicx}
\usepackage{graphics}
\usepackage{color}
\usepackage{cite}
\usepackage[leftbars]{changebar}
\newcommand{\vv}{{\mathbf v}}
\newcommand{\YY}{{\mathbf Y}}
\newcommand{\JJ}{{\mathbf J}}

\newcommand{\DD}{{\mathbf D}}
\newcommand{\KK}{{\mathbf K}}

\newcommand{\MM}{{\mathbf M}}
\newcommand{\CC}{{\mathbf C}}
\newcommand{\ww}{{\mathbf w}}
\newcommand{\hh}{{\mathbf h}}
\newcommand{\boldg}{{\mathbf g}}
\newcommand{\bsigma}{{\bm \sigma}}
\newcommand{\uu}{{\mathbf u}}
\newcommand{\uuu}{\underline{\mathbf u}}
\newcommand{\uFF}{\underline{\mathbf F}}
\newcommand{\uF}{\underline{F}}
\newcommand{\up}{\underline{p}}
\newcommand{\uww}{\underline{\mathbf w}}

\newcommand{\nn}{{\mathbf n}}
\newcommand{\ooo}{{\bm 0}}
\newcommand{\xx}{{\mathbf x}}

\newcommand{\ff}{{\mathbf f}}
\newcommand{\cA}{{\mathcal A}}

\newcommand{\cMte}{{\mathcal M}_t^e}
\newcommand{\cMoe}{{\mathcal M}_0^e}

\newcommand{\cD}{{\mathcal D}}
\newcommand{\cE}{{\mathcal E}}

\newcommand{\cH}{{\mathcal H}}
\newcommand{\cF}{{\mathcal F}}
\newcommand{\cB}{{\mathcal B}}
\newcommand{\cC}{{\mathcal C}}

\newcommand{\dd}{\text{d}}

\newcommand{\bxi}{\bm \xi}
\newcommand{\piijk}{\pi_{i,j,k}}
\newcommand{\vpiijk}{\varpi_{i,j,k}}
\newcommand{\bzeta}{\bm \zeta}

\newcommand{\sume}{\sum_{e=1}^E}

\newcommand{\JAt}{J_{\cA_t}}
\newcommand{\cAt}{\cA_t}
\newcommand{\parot}{\partial \Omega_t}
\newcommand{\parotcD}{\parot^\cD}
\newcommand{\paroo}{\partial \Omega_0}
\newcommand{\parote}{\partial \Omega_t^e}

\newcommand{\intot}[1]{\int_{\Omega_t} #1\,\dd \Omega}
\newcommand{\intote}[1]{\int_{\Omega_t^e} #1\,\dd \Omega}
\newcommand{\intdot}[1]{\int_{\parot^\bsigma} #1\,\dd \partial \Omega}
\newcommand{\intdott}[1]{\int_{\parot} #1\,\dd \partial \Omega}
\newcommand{\intdotte}[1]{\int_{\parot^e} #1\,\dd \partial \Omega}
\newcommand{\intdotD}[1]{\int_{\parotcD} #1\,\dd \partial \Omega}
\newcommand{\intdote}[1]{\int_{\parot^{e,\bsigma}} #1\,\dd \partial \Omega}
\newcommand{\ot}{\Omega_t}
\newcommand{\ote}{\Omega_t^e}
\newcommand{\otI}{\Omega_t \times I}
\newcommand{\oo}{\Omega_0}
\newcommand{\ooe}{\Omega_0^e}
\newcommand{\ho}{\hat{\Omega}}

\newcommand{\nx}[1]{{\bm \nabla}_{\xx}\cdot #1}

\newcommand{\nabx}[1]{{\bm \nabla}_\xx #1}
\newcommand{\Deltax}[1]{{\bm \Delta}_\xx #1}
\newcommand{\ddtx}[1]{\frac{\partial #1}{\partial t}{\biggr \vert}_{\xx} }
\newcommand{\ddtY}[1]{\frac{\partial #1}{\partial t}{\biggr \vert}_{\YY} }
\newcommand{\ddtYsmall}[1]{\partial #1/\partial t{\vert}_{\YY} }
\newcommand{\sddtY}[1]{{\partial #1}/{\partial t}{\vert}_{\YY} }
\newcommand{\sddtx}[1]{{\partial #1}/{\partial t}{\vert}_{\xx} }

\newcommand{\ddp}[2]{\frac{\partial #1}{\partial #2}}
\newcommand{\Rd}{\mathbb{R}^d}
\newcommand{\PP}{\mathbb{P}}

\newcommand{\Dxu}{{\DD_\xx (\uu )}}
\newcommand{\DDt}[1]{\frac{\dd #1}{\dd t}}
\newcommand{\conv}[1]{#1 \cdot \nabx #1}
\newcommand{\hnn}{\hat{\nn}}
\newcommand{\vhr}{(\hat{\vv} \circ \cAt^{-1})}
\newcommand{\vhrN}{(\hat{\vv}_N \circ \cAt^{-1})}
\newcommand{\vh}{\hat{\vv}}
\journal{\it App. Num. Math.}
\pubyear{2008}
\volume{In Press}
\begin{document}
\begin{frontmatter}
\title{Solution of moving-boundary problems by the spectral element method}
\author[EPFL]{Nicolas Bodard\thanksref{FNRS}},
\thanks[FNRS]{Supported by a Swiss National Science Foundation Grant No. 200020--101707}
\author[EPFL]{Roland Bouffanais\corauthref{cor}\thanksref{FNRS}},
\corauth[cor]{Corresponding author. Named appear in alphabetical order}
\ead{roland.bouffanais@epfl.ch}
\author[EPFL]{Michel O. Deville}
\address[EPFL]{Laboratory of Computational Engineering,\\ \'Ecole Polytechnique F\'ed\'erale de Lausanne,\\ STI -- ISE -- LIN, Station 9,\\ CH--1015 Lausanne, Switzerland}
\begin{abstract}
This paper describes a novel numerical model aiming at solving moving-boundary problems such as free-surface flows or fluid-structure interaction. This model uses a moving-grid technique to solve the Navier--Stokes equations expressed in the arbitrary Lagrangian-Eulerian kinematics. The discretization in space is based on the spectral element method. The coupling of the fluid equations and the moving-grid equations is essentially done through the conditions on the moving boundaries. Two- and three-dimensional simulations are presented: translation and rotation of a cylinder in a fluid, and large-amplitude sloshing in a rectangular tank. The accuracy and robustness of the present numerical model is studied and discussed.
\end{abstract}
\end{frontmatter}

\section{Introduction}
With the advent of powerful computational resources like clusters of PCs or parallel computers the numericists are able to address  more challenging problems involving multi-physics and multi-scale approaches. These problems cover a large spectrum of scientific and engineering applications. However, in this paper, for the sake of conciseness, we will restrict our attention to two specific problems, namely: free-surface flows and fluid-structure interaction.

Free-surface flows occur in many industrial applications: coating flows, vertical drawing of viscous fluids, jets, die flows, etc, and in environmental flows: ocean waves, off-shore engineering, coastal habitat and management, to name a few. Two review articles have been published in recent years and report the state-of-the-art of the field \cite{scardovelli99:_direc,tsai96:_comput}. It can be observed that free-surface flows have been tackled by direct numerical simulation at low and moderate Reynolds numbers. This reality is essentially due to the nonlinear characters of the flow. On top of the nonlinearity associated to the Navier--Stokes equations themselves, here we deal with a complicated geometry which is changing in time and which is part of the solution itself. This accumulation of difficulties calls for elaborate algorithms and numerical techniques.

Fluid-structure interaction has been recognized for a long time as a real challenge. Indeed, this interaction is present in engineering problems like turbo-machinery, aerospace applications: buffeting, acoustics, and also in biomedical flows like blood flow in the coronary arteries. Fluid-structure interaction is also encountered in the field of vortex-induced vibrations having many important marine applications (e.g related to oil exploration, cable dynamics, etc.). It is only at the present time that this type of interaction for three-dimensional cases appears to be feasible as the necessary computing power becomes available. On one hand, the computational fluid dynamics (CFD) codes integrate the full steady state or transient Navier--Stokes equations which govern the dynamics of a viscous Newtonian fluid. They mostly use finite volume or finite element approximations. On the other hand, the computational solid mechanics (CSM) codes integrate the dynamics of various solid models, incorporating for example, classical infinitesimal linear elasticity, nonlinear finite elasticity with large deformations, plasticity, visco-elasticity, etc. These problems are also highly nonlinear with respect to the complicated geometries at hand. The combination of the nonlinearities of the mathematical models for the constitutive relations and for the geometrical behaviour has called for a robust approach able to deal with all the complexities and intricacies. The finite element method (FEM) with the isoparametric elements has emerged as the leading technology and methodology in CSM.

In the present paper, the methodological framework is the same for the fluid and the solid parts and rests upon the spectral element method \cite{patera86:_spect,maday89:_spect_navier_stokes,deville02:_high,roenquist91:_spect_navier}. With this choice the space discretization is similar for both problems. As in free-surface flows and fluid-structure interaction the geometry is deforming and moving, it is needed to use the arbitrary Lagrangian--Eulerian (ALE) formulation \cite{hirt74:_arbit_lagran_euler,donea04:_arbit_lagran_euler_method,donea83:_arbit_lagran_euler,formaggia99:_arbit_lagran_euler}. This formulation allows to treat the full geometrical problem with respect to a reference configuration that is arbitrarily chosen. A mapping is introduced to ease the description of the current configuration with respect to a reference configuration. This process leads to an ALE velocity which will be related to a grid velocity.

 In Section 2, the mathematical models will be presented with the associated weak formulations in the ALE context. Section 3 will be devoted to space and time discretizations. Section 4 will describe the numerical algorithms for the moving-grid technique. Section 5 will present numerical results and the final section will draw some conclusions.

\section{Mathematical model}
\label{sec:math-model}
A moving boundary-fitted grid technique has been chosen to simulate the unsteady part of the boundary in our computations. In the particular cases dealt with in this paper, the unsteady part of the boundary can be either the free surface in case of free-surface flows, or for fluid-structure interaction problems, the interface between the fluid and the structure. This choice of a surface-tracking technique is primarily based on accuracy requirements. With this group of techniques, the grid is configured to conform to the shape of the interface, and thus adapts continuously---at each time step---to it and therefore provides an accurate description of the moving boundary to express the related kinematic and/or dynamic boundary conditions.

The moving-boundary incompressible Newtonian fluid flows considered in this paper, are governed by the Navier--Stokes equations comprising the momentum equation and the divergence-free condition. In the ALE formulation, a mixed kinematic description is employed: a Lagrangian description of the moving boundary, an Eulerian description of the fixed domain boundaries and a mixed description of the internal fluid domain.

\subsection{The ALE kinematic framework}
\label{subsec:ALE}
This section will introduce the notations used in this paper to define the variables and frames of reference related to the ALE formulation. The notations adopted hereafter are borrowed from \cite{nobile01:_numer_haemod,formaggia99:_arbit_lagran_euler}. We denote by $\ot$ the fluid domain subset of $\Rd$ with $d=2,3$ the space dimension, the subscript $t$ referring to the time $t$ as the fluid domain is changing when its boundaries are moving. Let us denote by $\oo$ a reference configuration---for instance the domain configuration at initial time $t=t_0$. The system evolution is studied in the time interval $I=[t_0,T]$. The position of a point in the current fluid domain $\ot$ is denoted by $\xx$---Eulerian coordinate---and in the reference frame $\oo$ by $\YY$---ALE coordinate. Let $\cAt$ be a family of mappings, which at each $t\in I$ associates a point $\YY \in \oo$ to a point $\xx \in \ot$:
\begin{equation} 
\cAt\ : \oo \subset \Rd \rightarrow \ot \subset \Rd, \qquad \xx(\YY,t)=\cAt (\YY).
\end{equation} 
$\cAt$ is assumed to be continuous and invertible on $\overline{\Omega}_0$ and differentiable almost everywhere in $I$. The inverse of the mapping $\cAt$ is also continuous on $\overline{\Omega}_0$.

The Jacobian matrix of the ALE mapping $\cAt$ is given by
\begin{equation} 
\JJ_{\cAt} = \ddp{\xx}{\YY},
\end{equation} 
and its determinant $\JAt = \text{det}\, \JJ_{\cAt}$ is the Jacobian of the mapping characterizing the metrics of $\ot$ generated from the one of $\oo$. The Euler expansion formula gives the relationship between the Jacobian of the mapping $\cAt$ and the divergence of the ALE velocity $\ww$:
\begin{equation}
  \label{eq:Euler_Expansion_Formula}
  \ddtY{\JAt}= \JAt \nx{\ww}, \qquad  \forall (\YY ,t) \in \oo \times I ,
\end{equation}
supplemented by the initial condition $\JAt=1$ for $t=t_0$. In real computations, $\ww$ will be associated to the mesh velocity. The hyperbolic partial differential equation \eqref{eq:Euler_Expansion_Formula} highlights the important role played by the divergence of the mesh velocity in the time evolution of the mapping $\cAt$. This particular point is emphasized in Section \ref{sec:moving-grid}, where one of the mesh-update techniques used in our simulations, enforces a divergence-free condition for $\ww$ resulting in a constant in time Jacobian $\JAt$ \cite{bouffanais05:_mesh_updat_techn_free_surfac}. Furthermore Eq. \eqref{eq:Euler_Expansion_Formula} constitutes the evolution law for $\JAt$ as in our formulation the ALE mesh velocity is calculated based on the essential boundary conditions of our problem in $\ot$, thereby defining the location of the grid nodes and the value of $\cAt$. 

Considering a time-dependent scalar field $f$ defined  on $\ot \times I$, the notation $\sddtY{f}$ refers to the time derivative in the ALE frame, or in short the ALE time derivative expressed in Eulerian coordinates as opposed to the regular time derivative in Eulerian coordinates and denoted by $\sddtx{f}$. Finally, the ALE mesh velocity $\ww$ is defined as
\begin{equation}
  \label{eq:w}
  \ww (\xx, t) = \ddtY{\cAt}.
\end{equation}
It is worth noting that a standard application of the chain rule to the time derivative gives
\begin{equation}
  \label{eq:time_derivative}
  \ddtY{f}= \ddtx{f} + \ww \cdot \nabx{f}.
\end{equation}
The symbol $\nabx{}$ indicates the gradient operation applied to the scalar field $f$ with respect to the Eulerian coordinate $\xx$. If $\ww = \ooo$, the mesh is fixed, and we recover the Eulerian description where $\sddtY{}$ is the classical time derivative $\sddtx{}$. If $\ww=\uu$ where $\uu$ is the fluid velocity field, we obtain the Lagrangian description and $\sddtY{}$ is the particle derivative. Eq. \eqref{eq:time_derivative} allows to generalize the Reynolds transport theorem for a time-dependent volume integral of a scalar field
\begin{equation}
  \label{eq:Generalized_RT}
 \DDt{} \left(  \intot{f}   \right) = \intot{\left( \ddtY{f} + f \nx{\ww} \right)}.
\end{equation}

\subsection{The strong ALE formulation for the Navier--Stokes equations}
\label{subsec:strong}
The governing equations of our moving-boundary problem in the ALE kinematic description, for an incompressible Newtonian fluid flow occupying a fluid domain $\ot$ whose boundary $\parot$ is evolving with time, are the Navier--Stokes equations which in strong form and in the Eulerian kinematic description read
\begin{align}
  \label{eq:NS_Strong_Eulerian}
  \ddtx{\uu} + \conv{\uu} &= -\nabx{p} + 2\nu \nx{\Dxu}  + \ff, & & \forall (\xx, t) \in \otI,\\
  \label{eq:Divergence_Free}
  \nx{\uu} &= 0,& & \forall (\xx, t) \in \otI,
\end{align}
where $\uu$ is the velocity field, $p$ the pressure field (normalized by the constant fluid density $\rho$ and relative to zero ambient), $\nu$ the kinematic viscosity of the fluid, $\Dxu = \frac{1}{2} ( \nabx{\uu} + \nabx{\uu}^T)$ the rate-of-deformation tensor and $\ff$ the body force per unit mass, with the superscript $T$ indicating the transpose. Eq. \eqref{eq:NS_Strong_Eulerian} expresses the conservation of momentum and the divergence-free condition \eqref{eq:Divergence_Free} is the continuity equation in its simplified form for an incompressible flow. Equations \eqref{eq:NS_Strong_Eulerian}--\eqref{eq:Divergence_Free} are valid in the internal fluid domain $\ot$, and have to be supplied with boundary conditions on the boundary $\parot$ and the problem being unsteady, an initial condition is also required. The initial velocity field is chosen as
\begin{equation}
  \label{eq:Initial_Velocity}
  \uu(\xx,t=t_0)=\uu^0(\xx), \qquad \forall \xx \in \ot, \quad \text{with } \ot = \Omega_{t_0} = \oo,
\end{equation}
such that $\nx{\uu^0}=0$.
Let the boundary $\parot$ be split into two non-overlapping parts $\parot = \parotcD \cup \parot^\bsigma$. In the sequel we will consider the two following types of boundary conditions
\begin{align}
  \label{eq:KBC}
  \uu &= \boldg(t), & &\text{on }\parotcD \text{ and }\forall t \in I,\\
  \label{eq:DBC}
  \bsigma \cdot \hnn = -p \textbf{I} \cdot \hnn +2 \nu \Dxu\cdot \hnn &= \hh(t), & &\text{on }\parot^\bsigma \text{ and }\forall t \in I,
\end{align}
where $\bsigma$ is the stress tensor, $\textbf{I}$ the identity tensor and $\hnn$ the local unit outward normal vector to $\parot$. Eq. \eqref{eq:KBC} is an essential boundary condition on $\parotcD$ of Dirichlet type. In the cases of free-surface flows and fluid-structure interaction problems, which are of particular interest for us, $\parotcD$ corresponds to fixed or prescribed moving solid walls where a no-slip condition has to be satisfied. Eq. \eqref{eq:DBC} is a natural boundary condition on $\parot^\bsigma$. For free-surface flows and fluid-structure interactions, $\parot^\bsigma$ represents prescribed inflow and/or outflow depending on the situations considered, but primarily the free surface itself or the interface between the fluid and the structure, where a mechanical equilibrium has to be enforced. Therefore \eqref{eq:DBC} is a dynamic boundary condition (DBC) expressing the continuity of the normal stress at the moving boundary. If free-surface flows are envisaged and if the surface tension is neglected, $\hh = -p_0 \hnn$ where $p_0$ is the pressure of the surrounding fluid, taken as zero in the sequel.

Using Eq. \eqref{eq:time_derivative}, we can recast the strong form of the conservation of the momentum of the Navier--Stokes equations in the ALE frame
\begin{equation}
  \label{eq:NS_Strong_ALE}
  \ddtY{\uu} + (\uu-\ww) \cdot \nabx{\uu} = -\nabx{p} + 2\nu  \nx{\Dxu}  + \ff,\qquad \forall (\xx, t) \in \otI,
\end{equation}
the divergence-free condition \eqref{eq:Divergence_Free}, the initial and boundary conditions \eqref{eq:Initial_Velocity}--\eqref{eq:DBC} remaining unchanged in the ALE kinematic description. Indeed, boundary conditions are related to the problem and not to the kinematic description employed, be it Eulerian, Lagrangian or arbitrary Lagrangian-Eulerian. Nevertheless the ALE mesh velocity $\ww$ is to a certain extent part of the unknown fields of the problem even though some freedom in moving the mesh makes the ALE technique so attractive. The details related to the treatment and the computation of the mesh velocity are presented in Section \ref{sec:moving-grid}.

\subsection{The weak ALE formulation for the moving-boundary problem governed by the Navier--Stokes equations}
\label{subsec:weak}
Based on the strong formulation of the moving-boundary problem described in Section \ref{subsec:strong}, one can derive the more appropriate weak transient ALE formulation. In a standard approach, first are introduced the spaces of test and trial functions used to express the initial problem in its weak form based on the reference configuration $\oo$. We may note that the spaces of test and trial functions for the pressure are identical as no essential Dirichlet condition is being imposed on this field. This space is the space of functions that are square Lebesgue-integrable on the domain $\ot$ and is denoted by $L^2(\ot )$. In general the velocity does not necessarily vanish on the domain boundary; in our particular case the existence of a non-homogeneous essential Dirichlet boundary condition on $\parotcD$ leads us to consider different spaces for the test and trial functions for the velocity field. The solution for the velocity $\uu$, of the problem \eqref{eq:NS_Strong_Eulerian}--\eqref{eq:DBC} will be searched for directly in the Sobolev space of trial functions $H^1_{\cD}(\ot)^d$ defined as follows
\begin{equation}
H^1_{\cD}(\ot)^d = \{\uu \in L^2 (\ot)^d, \ \ \nabx{u_i}\in L^2 (\ot)^d\ \textrm{with }i=1,\dots, d ,\ \  \uu_{{\vert}_{\parotcD}}=\boldg \},
\end{equation}
and corresponding to the current configuration $\ot$. The reference configuration $\oo$ will be used to build the velocity test functions $\vh$, which will be taken in the space $H^1_{0,\cD}(\oo)^d$ with
\begin{equation}
H^1_{0,\cD}(\oo) = \{\hat{v} \in L^2 (\oo), \ \ \nabx{\hat{v}}\in L^2 (\oo)^d,\ \  \hat{v}_{{\vert}_{\paroo^\cD}}=0 \},
\end{equation}
to satisfy a homogenous Dirichlet condition on $\paroo^\cD$. Subsequently, the dynamics of the test functions on the configuration $\ot$ is obtained by means of the existing inverse of the mapping $\cAt$. Therefore the velocity test functions on the configuration at time $t$ are the set of functions $\vhr$ with $\vh$ belonging to $H^1_{0,\cD}(\oo)^d$. In the sequel, the notation $\vhr$ is kept in order to emphasize two key points. First, the treatment of the weak form of the time derivative $\ddtYsmall{\uu}$ in Eq. \eqref{eq:NS_Strong_ALE} is based on the essential property that $\vh$ is not time dependent and consequently $\ddtYsmall{\vh}=\textbf{0}$. Second, such formulation highlights the path to follow when practically implementing the weak ALE formulation. Indeed, the time dependency is fully incorporated in the inverse mapping $\cAt^{-1}$ and the functions $\vh$ remains the same as the ones used in fixed-grid problems. Moreover, in a more general framework where $\cAt$ is still invertible but less regular, this formulation still holds and one only needs to care for the regularity of the functions $\vh$ and not of the functions $\vhr$. With the notations and spaces introduced, the weak transient ALE formulation reads:\\
\textit{Find $(\uu(t),p(t))\in H^1_{\cD}(\ot)^d \times L^2 (\ot)$ such that for almost every $t\geq t_0$}
\begin{align} \label{WNS}
\DDt{} \intot{\vhr \cdot \uu}+ \intot{\vhr \cdot \nx{[\uu\uu - \uu \ww]}}
&=  \nonumber  \\
\intot{(p \nx{\vhr} -2 \nu {\mathbf D}_\xx (\uu) : \nabx{\vhr} )}&\\
+\intot{\ff \cdot \vhr}+\intdot{\hh \cdot \vhr}, \ \ \ \ \ \ \ \ \ &\forall \hat{\vv} \in\, H_{0,\cD}^1(\oo)^d,\nonumber
\intertext{and}
-\intot{q \nx{\uu}}= 0, \ \ \ \ \ \ \ \ \forall q \in &\, L^2 (\ot).\label{WDF}
\end{align}
The above set of equations has to be intended in the sense of distribution in the interval $t>t_0$, therefore justifying the qualifier ``for almost every $t\geq t_0$'', see \cite{quarteroni94:_numer_approx_partial_differ_equat} for greater details. As expected the DBC \eqref{eq:DBC} appears `naturally' in the weak formulation above, corresponding to the last term on the right-hand side of \eqref{WNS} and being the only `surface term' as the spatial integration is limited to $\parot^\bsigma$. In addition, the DBC \eqref{eq:DBC} defines the reference pressure level and therefore no additional condition on the mean value of the pressure is required. Finally, it is well known (see \cite{quarteroni94:_numer_approx_partial_differ_equat} for instance) that a non-homogeneous Dirichlet boundary condition engenders a compatibility condition that the field $\uu$ has to satisfy. The origin of this condition is that, in order to be compatible with \eqref{eq:Divergence_Free} the boundary condition \eqref{eq:KBC} imposes
\begin{equation}
  \label{eq:Compatibility_Condition}
  \intdott{\uu \cdot \hnn} = \intdotD{\boldg(t) \cdot \hnn} + \intdot{\uu(t) \cdot \hnn}= 0 , \qquad \forall t \in I.
\end{equation}
Eq. \eqref{eq:Compatibility_Condition} is a consequence of \eqref{WDF} with $q=1$.

In order to ease the discretization of the continuous weak equations \eqref{WNS}--\eqref{WDF}, we introduce the following notations and bilinear forms, such as a scalar product defined by
\begin{align}
  \label{eq:1}
  (\uu,\vh) &:= \intot{\vhr \cdot \uu}, & & \forall \hat{\vv} \in\, H_{0,\cD}^1(\oo)^d,\\
  \intertext{a so-called `energy bilinear form'}
  \cA (\uu,\vh) &:= 2\nu \intot{{\mathbf D}_\xx (\uu) : \nabx{\vhr} },& & \forall \hat{\vv} \in\, H_{0,\cD}^1(\oo)^d,\\
  \intertext{a bilinear form related to the weak incompressibility constraint}
  \cB (\vh, p ) &:= - \intot{ p \nx{\vhr}}, & & \forall \hat{\vv} \in\, H_{0,\cD}^1(\oo)^d,\\
  \intertext{a trilinear form corresponding to the nonlinear convective term}
  \cC(\vh ; \uu,\ww) &:= \intot{\vhr \cdot \nx{[\uu\uu - \uu \ww]}}, & & \forall \hat{\vv} \in\, H_{0,\cD}^1(\oo)^d,\\
  \intertext{and finally two linear forms, the first one being related to the source term $\ff$}
  \cF (\vh ) &:= \intot{\ff \cdot \vhr}, & & \forall \hat{\vv} \in\, H_{0,\cD}^1(\oo)^d,\\
  \intertext{and the second one to the non-homogeneous natural boundary condition \eqref{eq:DBC} on $\parot^\bsigma$}
  \cH^\bsigma (\vh ) &:= \intdot{\hh \cdot \vhr}, & & \forall \hat{\vv} \in\, H_{0,\cD}^1(\oo)^d.
\end{align}
With these notations, the continuous weak ALE form of our moving-boundary Navier--Stokes problem can be recast as\\
\textit{Find $(\uu(t),p(t))\in H^1_{\cD}(\ot)^d \times L^2 (\ot)$ such that for almost every $t\geq t_0$}
\begin{align}
  \label{WNSBF}
  \DDt{} (\uu,\vh) + \cA(\uu,\vh) + \cB(\vh,p)+\cC(\vh; \uu,\ww) &= \cF(\vh) + \cH^\bsigma(\vh ), & & \hspace{-0.8ex}\forall \hat{\vv} \in\, H_{0,\cD}^1(\oo)^d, \\
  \cB(\uu , q) & = 0, & & \forall q \in L^2(\ot) . \label{WDFBF}
\end{align}

\section{Numerical technique and discretization}
\label{sec:num-tech-discr}
Moving-boundary problems, either free-surface or fluid-structure interaction, represent a real challenge not only for the mathematician but also for the numericists. As presented in Section \ref{sec:math-model}, the weak formulation of the problem \eqref{WNSBF}--\eqref{WDFBF} is an evidence of its complexity. In this section, particular emphasis will be put on the numerical space discretization of this arduous continuous problem. The general case with non-homogeneous natural and essential Dirichlet boundary conditions is dealt with, justifying the authors' choice of a very detailed presentation. The particular case of steady problems with non-homogeneous Neumann conditions and homogeneous Dirichlet boundary conditions was first formulated by Ho and Patera in \cite{ho91:_variat} in their study of free-surface flows dominated by inhomogeneous surface-tension effects. Furthermore, R{\o}nquist extended the previous work of Ho and Patera to the more general case of steady free-surface flow problems with non-homogeneous Neumann and Dirichlet boundary conditions \cite{roenquist96:_domain_decom_solver_three_dimen}. The specificities related to the treatment of transient problems is highlighted in the present paper, which to our knowledge is not available in the current literature.

\subsection{Spectral element discretization}
\label{sec:spectr-elem-discr}
The first step in the spectral element method (SEM) discretization consists in subdividing the fluid domain $\overline{\Omega}_t=\ot \cup \parot$ into $E$ non-overlapping elements $\{\ote\}_{e=1}^E$. In the sequel we will assume that this elemental subdivision is maintained for all values of $t$ in the interval $I$, therefore meaning that no re-meshing procedure is applied and leading to
\begin{equation}
  \label{eq:Elemental_Conservation}
  \ote = \cAt (\ooe), \qquad \text{for } e=1,\cdots , E, \qquad \qquad\forall t \in I.
\end{equation}
A re-meshing procedure for problem using SEM is presented in \cite{bouffanais05:_mesh_updat_techn_free_surfac} and can be used if needed. Each element $\ote$ involves a mesh constructed as a tensor product of one-dimensional grids. Although each space direction may be discretized independently of the others, without loss of generality we will consider only meshes obtained with the same number of nodes in each direction, denoted by $N+1$, corresponding to the dimension of the space of $N$th-order polynomials. To describe the discretization process accurately, we adopt the same procedure as in \cite{deville02:_high} and define the following spaces
\begin{equation}
  \label{eq:Space_Notations}
  X:=H^1_{0,\cD}(\oo)^d, \qquad Y:=H^1_\cD(\ot)^d,\qquad Z:=L^2(\ot).
\end{equation}

\subsection{Galerkin approximation}
\label{sec:galerk-appr}

We apply the Galerkin approximation to our Navier--Stokes problem in the ALE formulation in its weak form \eqref{WNSBF}--\eqref{WDFBF}, and therefore select finite dimensional polynomial subspaces $X_N$, $Y_N$ and  $Z_N$ to represent $X$, $Y$ and $Z$ respectively. A staggered-grid approach with element based on $\mathbb{P}_N-\mathbb{P}_{N-2}$ spaces for the velocity and pressure field respectively, allows to eliminate completely the spurious pressure modes \cite{maday92:_nimes_n_stokes}. In this context, the finite dimensional functional spaces are defined as
\begin{align}
  \label{eq:Spaces_Again1}
  X_N &:= X \cap \PP_{N,E}^d,\qquad  Y_N := Y \cap \PP_{N,E}^d,\qquad  Z_N := Z \cap \PP_{N-2,E},
\end{align}
with 
\begin{equation}
  \label{eq:PNE}
  \PP_{M,E} = \{ \phi | \phi \in L^2 (\ot) ; \phi |_{\ote} \text{ is a polynomial of degree }\leq M, \forall e=1,\cdots, E \}, 
\end{equation}
where the superscript $d$ in \eqref{eq:Spaces_Again1} reflects the fact that test and trial velocity functions are $d$-dimensional. With these notations the Galerkin approximation of \eqref{WNSBF}--\eqref{WDFBF} reads\\
{\it Find $(\uu_N(t),p_N(t)) \in Y_N \times Z_N$ such that for almost every $t\geq t_0$}
\begin{align}
  \label{eq:Galerkin_ALE_Momentum}
  \DDt{} \left( \uu_N , \vh_N  \right)+\cA (\uu_N,\vh_N) &+ \cB (\vh_N,p_N) +  & & \nonumber \\ 
  \cC(\vh_N;\uu_N,\ww_N) &=\cF_N (\vh_N)+\cH^\bsigma_N(\vh_N), & &\qquad &\forall \vh_N \in X_N, \\
  \label{eq:Galerkin_Divergence_Free}
  \cB(\uu_N,q_N)&=0, & &\qquad  &\forall q_N \in Z_N,
\end{align}
with 
\begin{align}
  \label{eq:Discrete_F}
  (\uu_N, \vh_N) &= \sume \intote{\uu_N \cdot \vhrN}, &\forall \vh_N \in X_N \\
  \cF_N(\vh_N) &= \sume \intote{\ff_N \cdot \vhrN}, &\forall \vh_N \in X_N\\
  \cH^\bsigma_N (\vh_N) &= \sume \intdote{\hh_N \cdot \vhrN}, &\forall \vh_N \in X_N
\end{align}
$\ff_N$ and $\ww_N$  being the projection of the fields $\ff$ and $\ww$ onto the finite dimensional space $\PP_{N,E}^d$.

The integrals within each of the spectral elements $\{{\ote}\}_{e=1}^E$ and on the boundaries $\{\parot^{e,\bsigma}\}_{e=1}^E$ are performed in a discrete manner using Gaussian quadrature rules. More specifically, all the terms in \eqref{eq:Galerkin_ALE_Momentum}--\eqref{eq:Galerkin_Divergence_Free} are integrated using a Gauss-Lobatto-Legendre (GLL) quadrature rule except for the two terms involving the bilinear form $\cB$ where functions discretized in $\PP_{N-2,E}$ appear. For these two terms, namely the pressure term and the divergence-free condition, a Gauss-Legendre (GL) quadrature rule is chosen. Therefore, the  $\PP_N-\PP_{N-2}$ Navier--Stokes problem in the ALE form is finally stated as\\
{\it Find $(\uu_N(t),p_N(t)) \in Y_N \times Z_N$ such that for almost every $t\geq t_0$}
\begin{align}
  \label{eq:Galerkin_ALE_Momentum_22}
  \DDt{} \left( \uu_N , \vh_N  \right)_{N,GLL}+ & \  \cA_{N,GLL} (\uu_N,\vh_N) + \cB_{N,GL} (\vh_N,p_N) + & & \nonumber \\
  \cC_{N,GLL}(\vh_N;\uu_N,\ww_N) &= \cF_{N,GLL} (\vh_N)+\cH^\bsigma_{N,GLL}(\vh_N), & &\forall \vh_N \in X_N, \\
  \label{eq:Galerkin_Divergence_Free_22}
  \cB_{N,GL}(\uu_N,q_N)&=0, & &\forall q_N \in Z_N.
\end{align}
To simplify the notations in the sequel, we will drop the subscript GLL and unless being explicitly specified, whenever an integration rule is required, the GLL one is implicitly being used.

\subsection{Semi-discrete Navier--Stokes moving-boundary problem in the ALE form}
\label{sec:semi-discrete-navier}
In order to formulate the semi-discrete version of our moving-boundary problem governed by the Navier--Stokes equations in the ALE form, we introduce the two tensor-product bases on the reference element $\ho:=[ -1,1]^d$ and for the sake of simplicity we will choose the same discretization order in each space direction $N$:
\begin{itemize}
\item the Gauss-Lobatto-Legendre Lagrangian interpolation basis of degree $N$
  \begin{align}
    \label{eq:piijk}
    \{ \piijk (\bxi )\}_{i,j,k=0}^N &:= \{ \pi_i(\xi_i )\}_{i=0}^N \otimes   \{ \pi_j(\xi_j )\}_{j=0}^N \otimes   \{ \pi_k(\xi_k )\}_{k=0}^N, & & & \forall \bxi \in \ho,
    \intertext{to expand the velocity field discretized on the $N$th-order GLL grid;
    \item the Gauss-Legendre Lagrangian interpolation basis of degree $N-2$
    }
    \label{eq:vpiijk}
    \{ \vpiijk (\bzeta )\}_{i,j,k=1}^{N-1}& := \{ \varpi_i(\zeta_i )\}_{i=1}^{N-1} \otimes   \{ \varpi_j(\zeta_j )\}_{j=1}^{N-1} \otimes   \{ \varpi_k(\zeta_k )\}_{k=1}^{N-1}, & & & \forall \bzeta \in \ho,
  \end{align}
  to expand the pressure field discretized on the GL grid of order $N-2$.
\end{itemize}
The expressions of the one-dimensional GLL and GL Lagrangian interpolant polynomials $\pi(\bxi)$ and $\varpi (\bzeta )$ appearing in \eqref{eq:piijk}--\eqref{eq:vpiijk} can be found in \cite{deville02:_high}. The polynomials $\{ \piijk (\bxi )\}_{i,j,k=0}^N$ and $\{ \vpiijk (\bzeta )\}_{i,j,k=1}^{N-1}$ will serve as bases for the functions in the spaces $X_N$, $Y_N$ and $Z_N$
\begin{align}
  \label{eq:basis_v}
  \uu_N(\xx(\bxi),t) &= \sum_{i,j,k=0}^N     \uu_{ijk}(t)\  \piijk (\bxi ), & \forall (\bxi,t) \in \ho\times I,\\
  \label{eq:basis_p}
  p_N (\xx(\bzeta),t)  &= \sum_{i,j,k=1}^{N-1} p_{ijk}(t)\    \vpiijk (\bzeta), & \forall (\bzeta,t) \in \ho \times I ,
\end{align}
where $\xx =\xx^e$ is the location of the point considered in the spectral element $\ote$, $\{\uu_{ijk}(t)\}_{i,j,k=0}^N$ the set of nodal values of the velocity field on the GLL grid of $\ote$ and $\{p_{ijk}(t)\}_{i,j,k=1}^{N-1}$ the set of nodal values of the pressure field on the GL grid of $\ote$. It is important to note that the time-dependency of the discretized velocity $\uu_N$ and pressure $p_N$ is not solely accounted by the time-dependent nodal values of these two fields. Indeed, due to the motion of the grid, the mapping between the position in the reference element $\ho$ and the spectral element $\ote$ at time $t$ is also time-dependent via the ALE mapping $\cAt$. If we note $\cMte$ (resp. $\cMoe$) the mapping from the reference element $\ho$ onto $\ote$ (resp. $\ooe$), we can write
\begin{align}
  \label{eq:mappings}
  \xx^e &= \cMte (\bxi), && \forall (\bxi,t) \in \ho\times I,\\
  \YY^e &= \cMoe (\bxi), && \forall (\bxi,t) \in \ho \times I,\\
  \xx^e &= \cAt (\YY^e), && \forall (\YY^e,t) \in \ooe \times I,
\end{align}
leading to following relationship between the different mappings
\begin{equation}
  \label{eq:mappings_again}
  \cMte = \cAt \circ \cMoe, \qquad \forall t \in I.
\end{equation}
Eq. \eqref{eq:mappings_again} shows that the second origin of the time-dependency of \eqref{eq:basis_v} and \eqref{eq:basis_p}, after the one due to the set of GLL and GL nodal values, is the moving-grid technique via the time-dependency of the ALE mapping $\cAt$.

Before embarking on the final process of semi-descritizing the equations for the moving-boundary problem, a last issue needs to be addressed: the treatment of the non-homogeneous Dirichlet boundary condition \eqref{eq:KBC} on $\parotcD$. First of all and as mentioned earlier, the non-homogeneity of \eqref{eq:KBC} leads to different spaces for the trial and test functions for the velocity field, $X_N$ and $Y_N$ respectively. Therefore the basis \eqref{eq:basis_v} developed for $X_N$ is not suitable for the solution $\uu_N(t)$ of the problem \eqref{eq:Galerkin_ALE_Momentum}--\eqref{eq:Galerkin_Divergence_Free} sought in $Y_N$. As presented earlier, the non-homogeneous Dirichlet boundary condition imposes to satisfy the compatibility condition \eqref{eq:Compatibility_Condition} whose discrete version reads
\begin{equation}
  \label{eq:Discrete_Compatibility_Condition}
  \sume \intdotte{\uu_N(t) \cdot \hnn} = 0, \qquad \forall t \in I.
\end{equation}
Let $\uu_{b,N}$ be a (piecewise) polynomial defined on the discrete boundary $\parote$ ($e=1,\dots,E$) and such that its nodal boundary values are equal to the boundary data $\boldg(t)$. In practice, the GLL Lagrangian interpolation bases defined on the element boundaries are used to expand $\uu_{b,N}$; however, in the rest of the inner domain these functions are zero. By construction, $\uu_{b,N}$ satisfies \eqref{eq:Discrete_Compatibility_Condition}. Setting $\uu_N=\uu_{0,N}+\uu_{b,N}$, the problem reduces to finding $\uu_{0,N}$ in the space $Y_{0,N}:=H^1_{0,\cD}(\ot)^d\cap \PP_{N,E}^d$. Therefore the problem \eqref{eq:Galerkin_ALE_Momentum}--\eqref{eq:Galerkin_Divergence_Free} can be reformulated as follows\\
{\it Find $(\uu_{0,N}(t),p_N(t)) \in Y_{0,N} \times Z_N$ such that for almost every $t\geq t_0$ }
\begin{align}
  \label{eq:Galerkin_ALE_Momentum_3}
  \DDt{} \left( \uu_{0,N} , \vh_N  \right)_{N}+& \ \cA_{N} (\uu_{0,N},\vh_N) + \cB_{N,GL} (\vh_N,p_N) + & & \nonumber \\
  \cC_{N}(\vh_N;\uu_{0,N},\ww_N)& = \cF_{N} (\vh_N)+\cH^\bsigma_{N}(\vh_N)+\cD_{1,N}(\vh_N,t) , &  &\forall \vh_N \in X_N, \\
  \label{eq:Galerkin_Divergence_Free_3}
  \cB_{N,GL}(\uu_{0,N},q_N)&=\cD_{2,N,GL}(q_N,t), & & \forall q_N \in Z_N,
\end{align}
with
\begin{align}
  \label{eq:D1}
  \cD_{1,N}(\vh_N,t) &=-  \DDt{} \left( \uu_{b,N}(t) , \vh_N  \right)_{N} & & \nonumber \\
  & -\cA_{N} (\uu_{b,N}(t),\vh_N) - \cC_{N}(\vh_N;\uu_{b,N}(t),\ww_N),& & \forall \vh_N \in X_N,\\
  \intertext{and}
  \label{eq:D12}
  \cD_{2,N,GL}(q_N,t) &=- \cB_{N,GL}(\uu_{b,N}(t),q_N),& &\forall q_N \in Z_N.
\end{align}
The two time-dependent terms $\cD_{1,N}$ et $\cD_{2,N,GL}$ appearing in \eqref{eq:Galerkin_ALE_Momentum_3} and \eqref{eq:Galerkin_Divergence_Free_3} are due to the non-homogeneity of the Dirichlet boundary condition. These values are related to the values of the discrete field $\uu_{b,N}(t)$, which as mentioned earlier, are obtained from the values of the field $\boldg (t)$ from \eqref{eq:KBC}.\\

We can now expand the trial velocity $\uu_{0,N}$ and the trial pressure $p_N$ onto the GLL--GL bases like in \eqref{eq:basis_v} and \eqref{eq:basis_p} respectively. In the sequel we will drop the subscript $0$ in $\uu_{0,N}$, no confusion being possible as the non-homogeneous Dirichlet boundary conditions is already accounted for in \eqref{eq:Galerkin_ALE_Momentum_3}--\eqref{eq:Galerkin_Divergence_Free_3}. The semi-discrete equations derived from \eqref{eq:Galerkin_ALE_Momentum_3}--\eqref{eq:Galerkin_Divergence_Free_3} are
\begin{align}
  \label{eq:SD_Momemtum}
  \DDt{}(\MM(t) \uuu(t))&= - \KK (t) \uuu(t) - \CC (\uuu(t),\uww(t),t) \uuu(t)+\DD^T(t) \up(t)+ \uFF_1(t),\\
  \label{eq:SD_Divergence_Free}
  -\DD(t) \uuu(t) &= \uF_2(t).
\end{align}
The matrices appearing in \eqref{eq:SD_Momemtum}--\eqref{eq:SD_Divergence_Free} are all time-dependent: $\MM$ is the mass matrix, $\KK$ the stiffness matrix, $\CC$ the discrete convective operator involving the velocity field $\uuu$ and the ALE mesh velocity $\uww$, $\DD^T$ the discrete gradient operator and $\DD$ the discrete divergence. $\uFF_1$ and $\uF_2$ are two vectors accounting for the presence of the body force $\ff$ and the time-dependent essential Dirichlet and natural non-homogeneous boundary conditions.

\subsection{Time discretization}
\label{subsec:time-dis}
The set of semi-discrete equations \eqref{eq:SD_Momemtum}--\eqref{eq:SD_Divergence_Free} is discretized in time using finite-difference schemes in a decoupled approach. The computation of the linear Helmholtz problem---corresponding to the energy bilinear form $\cA$ and the stiffness matrix $\KK$---is integrated based on an implicit backward differentiation formula of order 2, the nonlinear convective term---corresponding to the trilinear form $\cC$ and the matrix $\CC$---is integrated based on a relatively simple extrapolation method of order 2, introduced by Karniadakis et al. \cite{karniadakis91:_high_navier}.

The moving-grid treatment requires the semi-discrete equations \eqref{eq:SD_Momemtum}--\eqref{eq:SD_Divergence_Free} to be supplemented with an equation computing the mesh nodes update
\begin{equation}
  \label{eq:mesh-update}
  \DDt{\underline{\xx}} = \underline{\ww},
\end{equation}
with $\underline{\xx}$ being a vector containing the $d$-dimensional mesh nodes position at time $t$. The integration of Eq. \eqref{eq:mesh-update} necessitates the knowledge of the values of the mesh velocity provided by the moving-grid technique employed. Two particular techniques are presented in detail in Section \ref{sec:moving-grid}. The time-integration of Eq. \eqref{eq:mesh-update} is based on an explicit and conditionally stable Adams--Bashforth of order 3.

Lastly the treatment of the pressure relies on a generalized block LU decomposition with pressure correction \cite{perot93,perot95:_commen}.

The temporal order of the overall splitting scheme has proved to be equal to two for fixed-grid problems. The grid motion induces a limited reduction of the global temporal order, which is found to fall between 1.5 and 2 for the simulations presented in Section \ref{sec:num-sim-res}. The reasons for this reduction in the global order of the method is still not clearly understood.

\subsection{Specificities pertaining to free-surface flows and fluid-structure interaction}
\label{subsec:speci}
Up to this point, the treatment of our moving-boundary problem was kept to a level general enough to encompass both the free-surface flow and fluid-structure interaction problems. At this stage, it appears natural to provide the specificities pertaining to each of these two sub-problems.

These specificities lie primarily in the natural and essential Dirichlet boundary conditions imposed to the system. For free-surface flows with no surface-tension effects---either of normal or tangential type---and with no inflow nor outflow---closed system, both the natural and essential Dirichlet boundary conditions become homogeneous---$\boldg = \textbf{0}$ on $\parotcD$ and $\hh = \textbf{0}$ on $\parot^\bsigma$, leading to a drastic simplification of the problem. More precisely, both vectors $\uFF_1$ and $\uF_2$ vanish in the semi-discrete formulation \eqref{eq:SD_Momemtum}--\eqref{eq:SD_Divergence_Free} of the problem. For fluid-structure interaction problems, the natural boundary condition on the interface between the flow and the structure is provided by the dynamics of the structure, that can be evaluated by the SEM and the Newmark method \cite{bodard04:_fluid}. In the sequel, we will focus our attention on the flow problem for both of these two cases of interest.  

\section{Moving-grid techniques}
\label{sec:moving-grid}
When considering moving-boundary problems tackled in an interface-tracking and ALE frame, the moving boundary $\parot^\bsigma$ is treated in a Lagrangian way whereas the boundary $\parotcD$ which is fixed or subject to a prescribed motion $\boldg (t)$, is studied in an Eulerian frame. Accordingly, such method allows large-amplitude motions of the moving boundary, by generating a shape-conformed grid. Hence, an accurate and simple application of the boundary conditions on $\parot$ is easily accessible.

As a consequence of the ALE kinematics, the boundary conditions imposed on the mesh velocity $\ww$ read
\begin{align}
  \label{eq:W_KBC}
  \ww \cdot \hnn &= \uu \cdot \hnn, &   &\text{on } \parot^\bsigma \text{ and }\forall t \in I,\\
  \label{eq:W_KBC2}
  \ww &= \boldg(t), &  &  \text{on } \parotcD \text{ and }\forall t \in I.
\end{align}
Eq. \eqref{eq:W_KBC} is a kinematic boundary condition (KBC) on the moving boundary traducing that $\parot^\bsigma$ is a material surface with no transfer of fluid across it in the Lagrangian perspective. Eq. \eqref{eq:W_KBC2} expresses a kinematic boundary condition of no-slip type on the boundary of the domain which is not free to move. Given \eqref{eq:W_KBC}--\eqref{eq:W_KBC2}, it appears that the freedom left for the choice of $\ww$ lies in the values of this field in the internal fluid domain $\ot$ and also on the tangential values of $\ww$ on the moving boundary $\parot^\bsigma$. 

The computation of the mesh velocity $\ww$ in the internal fluid domain $\ot$ is the corner-stone of the moving-grid technique developed in the framework of the ALE formulation. The values of the mesh velocity being prescribed on the boundary $\parot$ as expressed by equations \eqref{eq:W_KBC}--\eqref{eq:W_KBC2}, the evaluation of $\ww$ in $\ot$ can be obtained as the solution of an elliptic equation:
\begin{align} \label{ELLIPTIC}
\cE_\xx\ww &= {\mathbf 0}, & & \forall (\xx, t) \in \otI,
\end{align} 
where $\cE_\xx$ represents any elliptic operator based on the Eulerian coordinates $\xx$. Such elliptic equation constitutes a classical choice for calculating the mesh velocity \cite{ho90:_legen}. Two types of elliptic equations are envisaged in this paper. The first elliptic operator used is a Laplacian operator ${\bm \Delta}_\xx$, and will be used extensively in the fluid-structure interaction simulations presented in Section \ref{sec:num-sim-res}. More details about the use of this specific operator for the computation of the mesh velocity can be found in \cite{bodard04:_fluid}. The second strategy relies on the assumption that the motion of the mesh nodes is equivalent to a steady Stokes flow, corresponding physically to an incompressible and elastic motion of the mesh. Therefore the problem for the mesh velocity becomes a boundary-value steady Stokes problem with the mesh velocity $\ww$ satisfying a divergence-free condition
\begin{align}
  \nx{\ww} &= 0 & &\forall (\xx, t) \in \otI. \label{SETLAST}
\end{align}
The justifications of this additional constraint imposed to the mesh velocity problem is presented in detail in \cite{bouffanais05:_mesh_updat_techn_free_surfac}. The free-surface flow simulations of a sloshing in a three-dimensional tank due to the gravity  presented in Section \ref{sec:standing-waves}, were carried out using this second strategy for $\ww$.

\section{Numerical simulations and results}
\label{sec:num-sim-res}
In this section we will present numerical results for three problems: the steady Stokes equations in curved subdomains, the motion of a cylinder in a square cavity and the sloshing in a three-dimensional tank. 
\subsection{Accuracy in curved domains}
\begin{figure}[h]
  \centering
  \includegraphics[width=12cm, keepaspectratio=true]{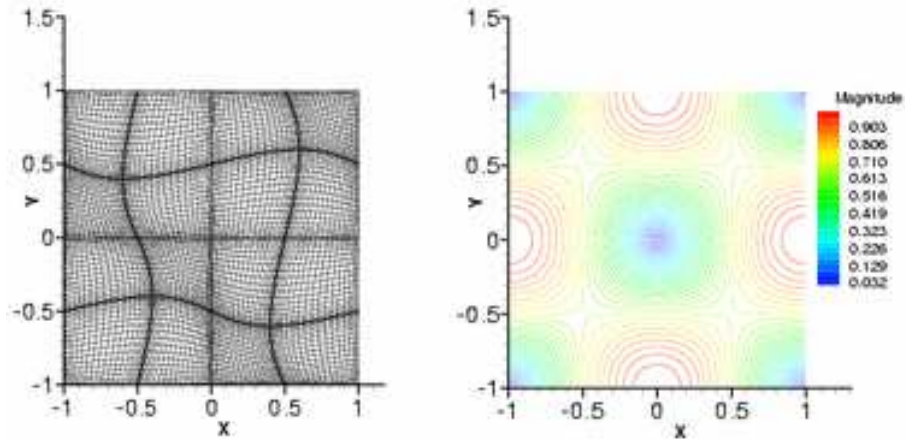}
  \caption{Square domain $\Omega$ with internally deformed subdomains with $E=4\times 4$ spectral elements and $N=20$ (left) and the velocity magnitude (right)}\label{fig1}
\end{figure}

 We want to check the error evolution  in the square domain $\Omega=[-1,1]^2$ decomposed in curved subdomains (elements).  
 To this aim, let us consider the steady Stokes equations  
 \begin{align}
   -\nabx{p} + \Deltax \uu + \ff &= \textbf{0}, & & \forall (\xx, t) \in \otI,\label{Sto1}\\
   \nx{\uu} &= 0,& & \forall (\xx, t) \in \otI.\label{Sto2}
 \end{align}
 The exact solution is given by 
 \begin{align}
   u_x&=-\cos\left(\pi x/2\right) \sin\left(\pi y/2\right),\\
   u_y&=\sin\left(\pi x/2\right) \cos\left(\pi y/2\right),\\
   p  &=-\pi \sin\left(\pi x/2\right) \sin\left(\pi y/2\right), 
 \end{align}
 when the body force term is chosen as
 \begin{equation}\label{ffunc}
     f_x=-\pi^{2} \cos\left(\pi x/2\right) \sin\left(\pi y/2\right) ,\qquad f_y=0.
 \end{equation}
 Instead of using a regular square grid composed of elements with edges parallel to the lines of the Cartesian axes, we performed the computation on the deformed mesh \cite{schneidesch92:_numer_cheby,gordon73:_const} displayed in Figure \ref{fig1} (left). Contours of the norm of the velocity field for the computed solution of problem \eqref{Sto1}--\eqref{Sto2}, are presented on Figure \ref{fig1} (right). Figure \ref{fig2} shows the evolution of the relative error in $H^1$-norm for the velocity and in $L^2$-norm for the pressure field, with respect to an increasing polynomial degree $N$, for two cases---$E=2\times2$ elements and $E=4\times4$ elements. The convergence is slower than the one obtained with a mesh divided in several regular square subdomains. First, we still achieve the exponential decrease of the relative error when the polynomial degree increases (which is typical of spectral or $p$-convergence \cite{deville02:_high}). Second, the convergence is faster when increasing the number of spectral elements $E$, as previously observed in \cite{schneidesch93:_cheby_navier} (which is equivalent to $h$-convergence in finite-element terminology \cite{deville02:_high}).

\begin{figure}[h]
  \centering
  \input{RelativeNorm_InternalCurved.pstex}
  \caption{Relative error in $H^1$-norm for the velocity and in $L^2$-norm for the pressure field}\label{fig2}
\end{figure}

\begin{figure}[h]
  \centering
  \includegraphics[width=12cm, keepaspectratio=true]{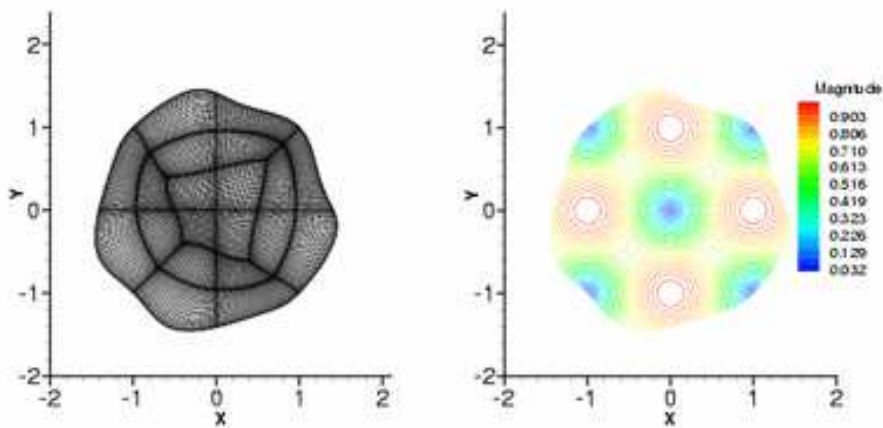}
  \caption{Curvy geometry for $E=20$ and $N=30$ (left) and the velocity magnitude (right)}\label{fig3}
\end{figure}

The same computation has been carried out with a geometry $\Omega'$ obtained by the transformation of coordinates of the unit square $\Omega=[-1,1]^2$ with sine functions (Fig. \ref{fig3})
\begin{align}\label{PotTrans}
    x'&=x+\alpha\sin\left(\pi x\right) \sin\left(\pi y\right),\\
    y'&=y+\alpha\sin\left(\pi y\right) \sin\left(\pi y\right),
\end{align}
with $(x,y)\in \Omega$, $(x',y')\in \Omega'$ and $\alpha=1/10$. Here, the deformation of the geometry not only involves the interior of the subdomains but also the domain boundaries.

\begin{figure}[h]
  \centering
  \input{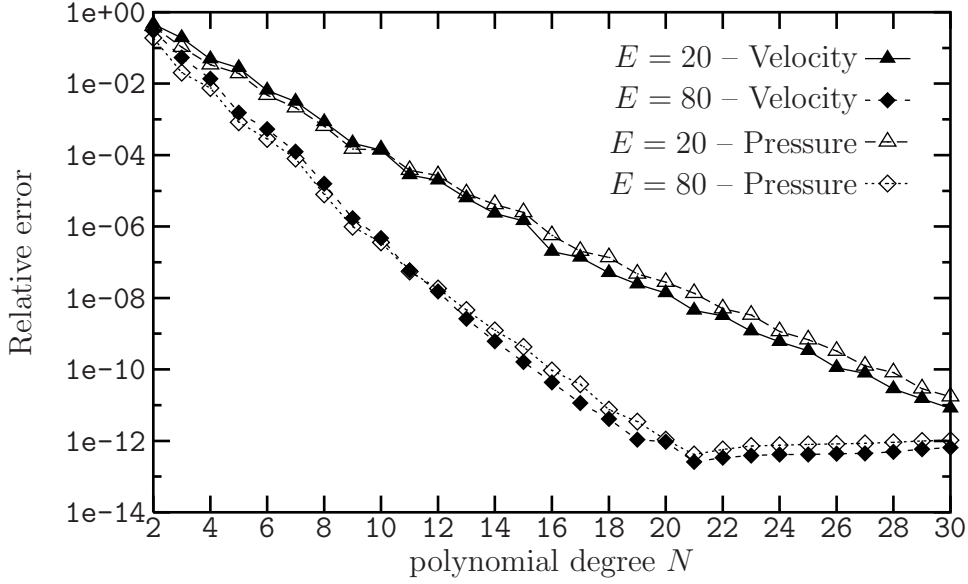}
  \caption{Relative error in $H^1$-norm for the velocity and in $L^2$-norm for the pressure field}\label{fig4}
\end{figure}

The remarks made for the first computation, corresponding to the square domain, are still relevant for this geometry. The same behavior of the convergence is obtained as one can observe on Figure \ref{fig4}. The deformation of the boundaries induces obviously a slower convergence in comparison with the square domain. Nevertheless, the important result is that the spectral convergence is maintained (Fig. \ref{fig4}) even with a deformation of the domain involving its boundaries, which is a mandatory feature when solving moving-boundary problems.

\subsection{Motion of a cylinder inside a square cavity}
\begin{figure}[h]
  \centering
  \input{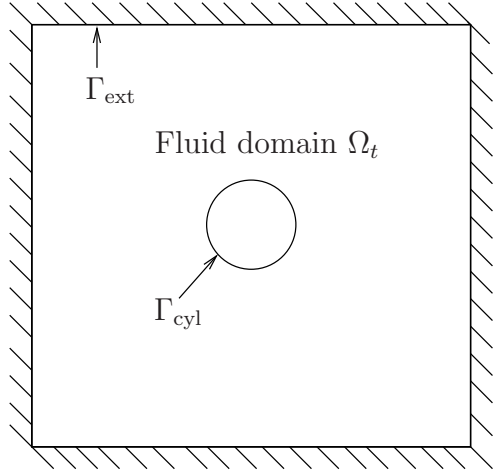}
  \caption{Geometry of the fluid domain with the immersed cylinder}\label{figMG}
\end{figure}

\begin{figure}[h]
  \centering
  \includegraphics[width=4.5cm, keepaspectratio=true]{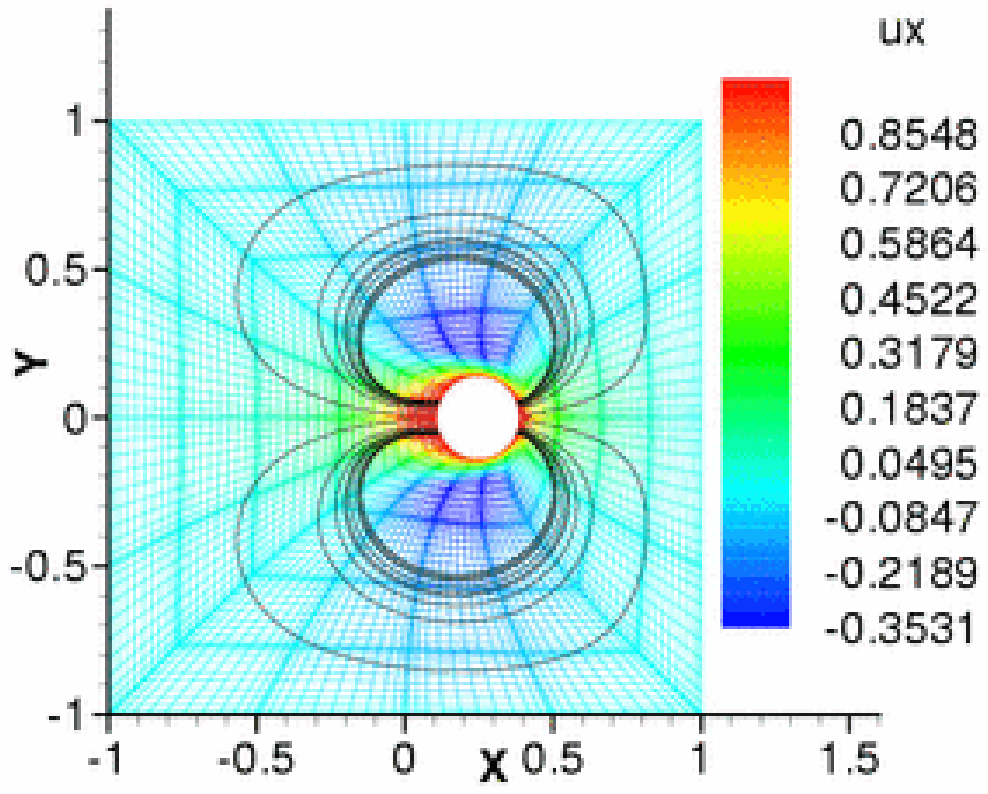}
  \includegraphics[width=4.5cm, keepaspectratio=true]{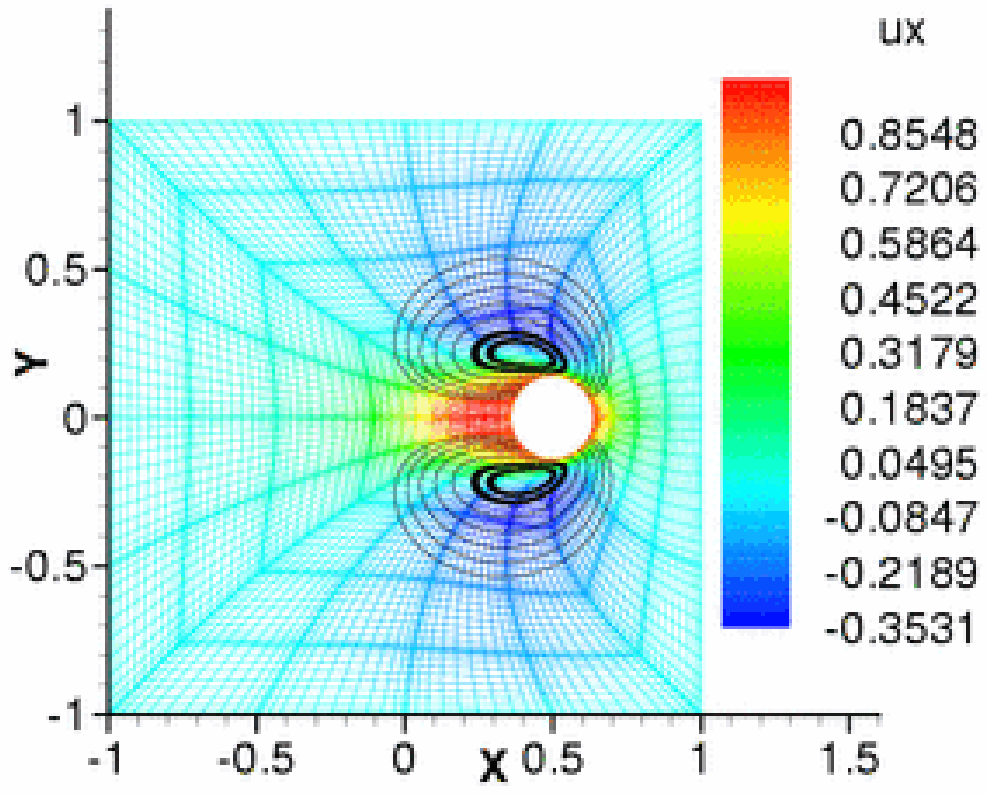}
  \includegraphics[width=4.5cm, keepaspectratio=true]{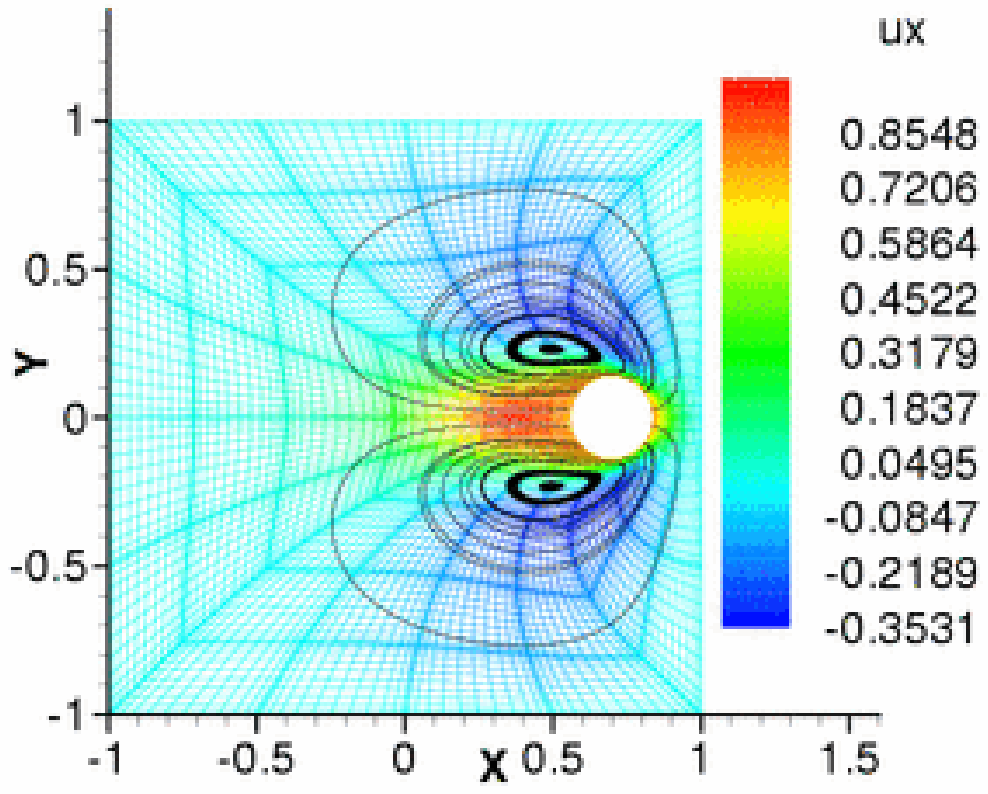}
  \caption{The velocity component $u_x$ and the corresponding streamlines (black solid lines) around a moving cylinder, $\textrm{Re}=100$, for $t=0.25$ (left), $t=0.5$ (center) and $t=0.7$ (right)}\label{fig5}
\end{figure}

We solve the Navier--Stokes equations \eqref{eq:NS_Strong_Eulerian}--\eqref{eq:Divergence_Free} in a two-dimensional square cavity. A schematic view of this cavity is given in Figure \ref{figMG} with fixed exterior walls. A circular cylinder is immersed into the fluid and is moving with a prescribed velocity. Two types of prescribed motions are studied. In the first case, we consider a cylinder in horizontal translation from the center of the cavity with a constant velocity. Denoting the boundary of the cylinder as $\Gamma_{\textrm{cyl}}$, we prescribe
\begin{align}\label{VelT}
    u_x\vert{_{\Gamma_{\textrm{cyl}}}}=w_x\vert{_{\Gamma_{\textrm{cyl}}}}&=1,\\
    u_y\vert{_{\Gamma_{\textrm{cyl}}}}=w_y\vert{_{\Gamma_{\textrm{cyl}}}}&=0.
\intertext{Denoting the exterior walls as $\Gamma_{\textrm{ext}}$, we have}
\label{VelExt}
\uu|_{\Gamma_{\textrm{ext}}}=\ww|_{\Gamma_{\textrm{ext}}}&=\textbf{0}.
\end{align}
We solve a time-dependent problem in order to study the evolution of the fluid motion caused by the translation of the cylinder in the square domain $\Omega=[-1,1]^2$. The Reynolds number based on a unit reference length and a unit reference velocity is $\textrm{Re}=1/\nu=100$. The diameter of the cylinder is $D=0.28$. The time step $\Delta t$ is fixed to $0.005$. The discretization uses a total number of elements equal to $E=64$ and the polynomial degree is $N=12$ in each of the two directions. We obtain an unsteady evolution of the fluid motion and we observe a deformation of the fluid mesh as pictured in Figure \ref{fig5}, where appears the flow configuration for $t=0.25$, 0.5 and 0.7. If we keep on moving the cylinder closer to the right wall, the mesh deformation becomes too large. We have also focused our attention on the evaluation of an artificial ``acceleration'' defined as $\|\uu_{n+1}-\uu_{n}\|_{L^{2}}/\Delta t$, where $\uu_{n}$ denotes the velocity field at the time level $n$. Figure \ref{fig7} displays the previous expression and the $L^{2}$-norm of the velocity. The ``acceleration'' does not vanish, which means the solution does not become steady-state. This can be expected since the cylinder is always in motion inside the cavity. Moreover the ``acceleration'' increases when the cylinder gets closer to the right wall.

\begin{figure}[h]
  \input{NormT.pstex}
  \caption{$\|\uu_{n+1}-\uu_{n}\|_{L^{2}}/\Delta t$ and $\|\textbf{u}\|_{L^{2}}$ versus simulation time $t$}\label{fig7}
\end{figure}

\begin{figure}[h]
  \centering
  \includegraphics[width=4.5cm, keepaspectratio=true]{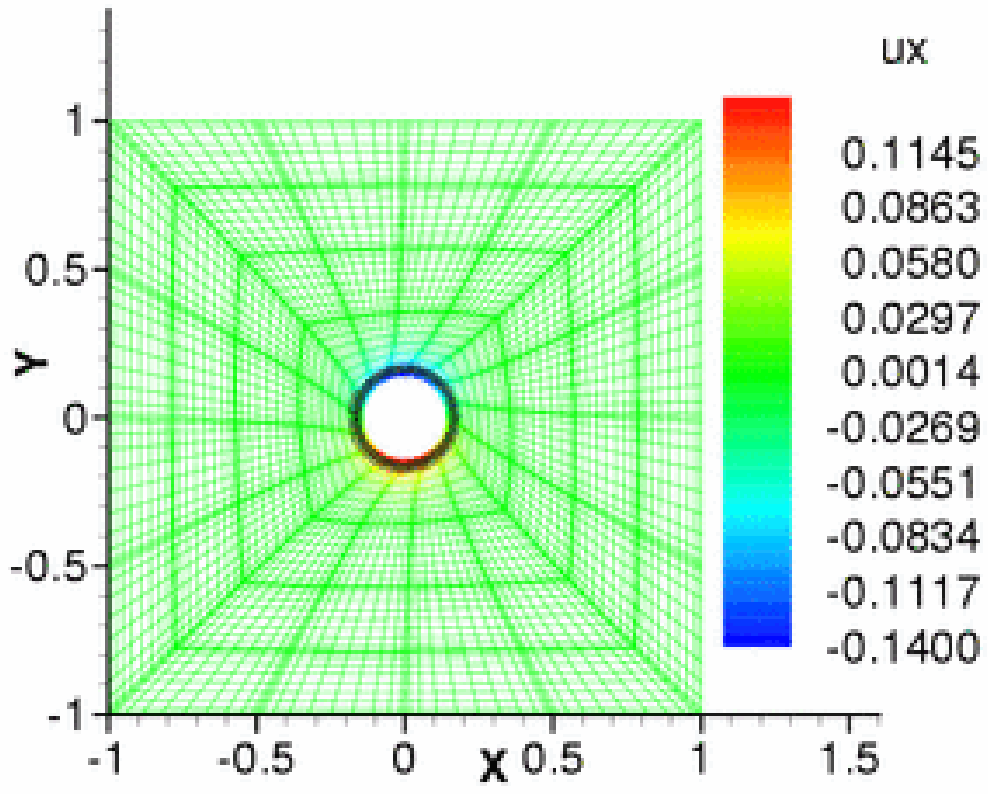}
  \includegraphics[width=4.5cm, keepaspectratio=true]{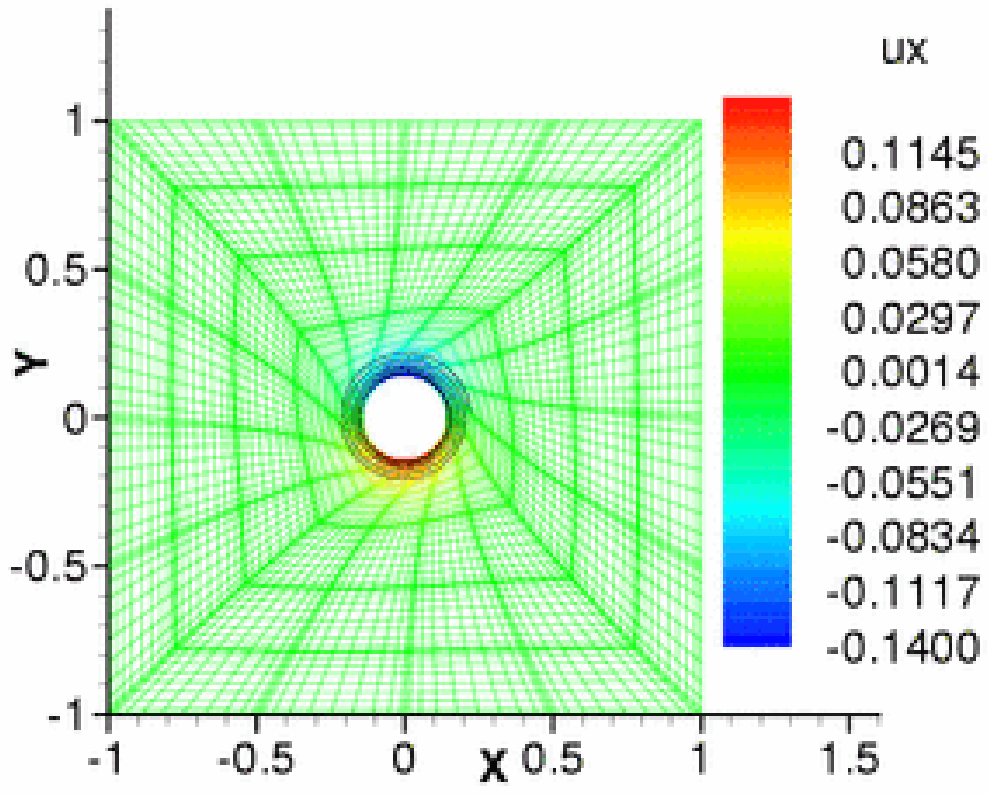}
  \includegraphics[width=4.5cm, keepaspectratio=true]{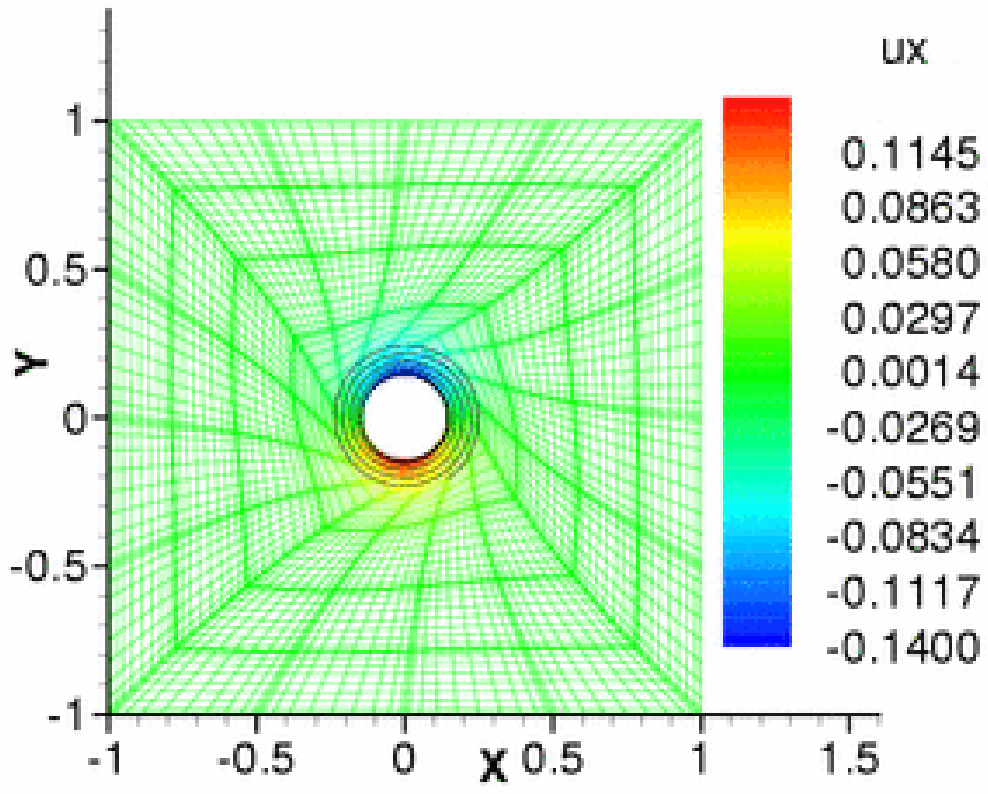}
  \caption{The velocity component $u_x$ and the corresponding streamlines (black solid lines) around a moving cylinder, $\textrm{Re}=100$, for $t=0.50$ (left), $t=1.25$ (center) and $t=2.0$ (right)}\label{fig8}
\end{figure}

In the second case, the cylinder at the center of the cavity is subject to a constant counter-clockwise angular rotation $\omega=1$ such that
\begin{align}\label{VelR}
    u_x|_{\Gamma_{\textrm{cyl}}}=w_x|_{\Gamma_{\textrm{cyl}}}&=-y,\\
    u_y|_{\Gamma_{\textrm{cyl}}}=w_y|_{\Gamma_{\textrm{cyl}}}&=x.
\end{align}
In Figure \ref{fig8}, we have maintained $E=64$ and changed the polynomial
degree to $N=10$. We exhibit the flow configuration for $t=0.5$, 1.25 and 2.0. Like in the previous example, we conclude the solution does not reach a steady state due to the motion of the cylinder (Fig. \ref{fig10}).
We have successfully tested these kinds of motion for large distortions of the fluid mesh.

\begin{figure}[h]
  \input{NormR.pstex}
  \caption{$\|\uu_{n+1}-\uu_{n}\|_{L^{2}}/\Delta t$ and $\|\textbf{u}\|_{L^{2}}$ versus simulation time $t$}\label{fig10}
\end{figure}

\subsection{Sloshing in a three-dimensional tank}
\label{sec:standing-waves}
\begin{figure}[h]
  \centering
  \input{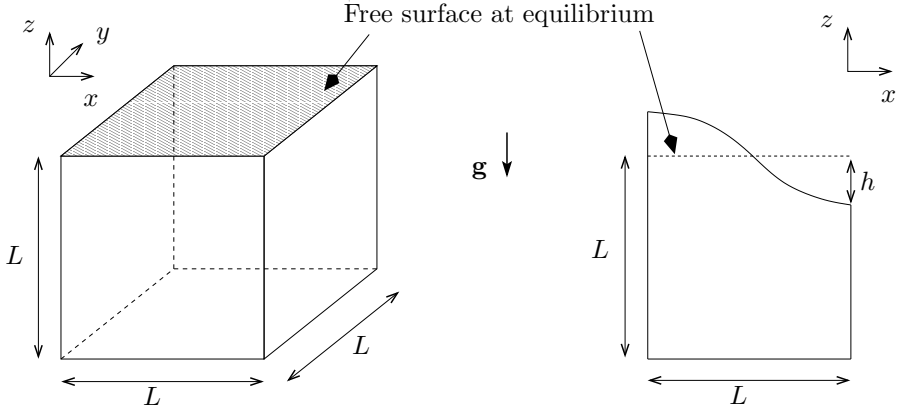}
  \caption{Geometry of the simulation set-up and rectangular tank}\label{fig:tank_geo}
\end{figure}

To show the adaptability of our numerical model based on a moving-grid technique in the ALE frame, the analysis of large-amplitude sloshing in a three-dimensional rectangular tank has been carried out. The tank has a square-base section of dimensions $L\times L$ and a schematic view of the geometry of the problem is presented in Figure \ref{fig:tank_geo}. The free-surface position is measured from the bottom of the tank: $z=L+h(x,y,t)$, where $h(x,y,t)$ is the free-surface elevation measured from its equilibrium position $z=L$. The initial shape of the free surface is varying only with $x$ and is given by the nonlinear theory for finite-amplitude standing waves
\begin{align}
  \label{eq:finite_amplitude_wave}
  h(x,t) = a(t) \cos (kx) \cos (\omega t)& \nonumber \\
 -\frac{ka^2(t)\cos (2 kx )}{2 \tanh (kL)} &\left\{  \cos^2(\omega t) + \frac{3 \cos (2\omega t) - \tanh^2 (kL)}{4 \sinh^2 (kL)} \right\},
\end{align}
where the wave number is $k=2\pi / \lambda$, the wave length $\lambda=2L$, the initial wave amplitude $a(t=0)=L/5$, and $\omega = \sqrt{gk \tanh ( kL ) }$ corresponding to the dispersion relation. For the sake of simplicity, we have taken $g=2\pi \lambda \tanh (kL)$ which leads to a period $T$ of the non-viscous and irrotational waves equal to one. Finally, the Reynolds number is based on the reference velocity $\sqrt{gL}$ and is expressed as $\textrm{Re}=L\sqrt{gL}/\nu$.

A no-slip condition is imposed to the velocity field at the bottom of the tank $z=0$ and free-slip conditions on the side walls $x=0$, $x=L$ and $y=0$, $y=L$, likewise for the mesh velocity field $\ww$ which corresponds to Eq. \eqref{eq:W_KBC2} in the context of this problem. The Dirichlet boundary condition on the free surface for the mesh velocity is given by $\ww=\uu$, which includes Eq. \eqref{eq:W_KBC} and also an additional condition on the tangential values of $\ww$. The initial velocity field is the irrotational solution at the maximum displacement of a standing wave---corresponding to zero for all velocity components. When starting the simulation, the top surface is set free and allowed to evolve in response to the dynamic and kinematic boundary conditions \eqref{eq:DBC} and \eqref{eq:W_KBC} respectively. The nonlinearity of this problem is introduced by both boundary conditions, through the shape of the free surface in \eqref{eq:W_KBC} and through the normal to the free surface in \eqref{eq:DBC}. The motion of the free surface physically corresponds to a transfer of energy between the potential energy---maximum at the initial time---and the kinetic energy, leading to a decaying oscillatory phenomenon.

Computations are performed with $E=3^3=27$ spectral elements and a polynomial degree $N=9$ in all three directions, leading to a mesh comprising $28^3$ nodes. The time step is taken equal to 0.001 and the simulation duration is 25 time units---based on the unit period $T$---or 25'000 iterations for 7 values of the Reynolds number $\textrm{Re}=50$, 100, 250, 500, 750, 1'000 and 1'500.

\begin{figure}[h]
  \centering
  \input{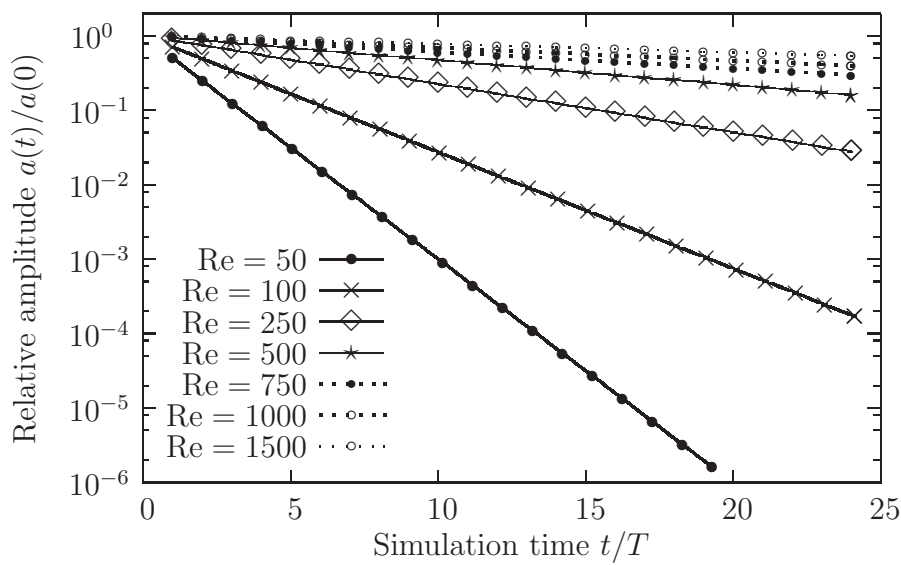}
  \caption{Relative amplitude $a(t)/a(0)$ against the simulation time $t/T$}\label{fig:wave_damping}
\end{figure}

\begin{figure}[h]
  \centering
\setlength{\unitlength}{0.240900pt}
\ifx\plotpoint\undefined\newsavebox{\plotpoint}\fi
\sbox{\plotpoint}{\rule[-0.200pt]{0.400pt}{0.400pt}}%
\begin{picture}(1500,900)(0,0)
\sbox{\plotpoint}{\rule[-0.200pt]{0.400pt}{0.400pt}}%
\put(281.0,123.0){\rule[-0.200pt]{4.818pt}{0.400pt}}
\put(261,123){\makebox(0,0)[r]{$10^{-6}$}}
\put(1419.0,123.0){\rule[-0.200pt]{4.818pt}{0.400pt}}
\put(281.0,158.0){\rule[-0.200pt]{2.409pt}{0.400pt}}
\put(1429.0,158.0){\rule[-0.200pt]{2.409pt}{0.400pt}}
\put(281.0,205.0){\rule[-0.200pt]{2.409pt}{0.400pt}}
\put(1429.0,205.0){\rule[-0.200pt]{2.409pt}{0.400pt}}
\put(281.0,229.0){\rule[-0.200pt]{2.409pt}{0.400pt}}
\put(1429.0,229.0){\rule[-0.200pt]{2.409pt}{0.400pt}}
\put(281.0,240.0){\rule[-0.200pt]{4.818pt}{0.400pt}}
\put(261,240){\makebox(0,0)[r]{$10^{-5}$}}
\put(1419.0,240.0){\rule[-0.200pt]{4.818pt}{0.400pt}}
\put(281.0,275.0){\rule[-0.200pt]{2.409pt}{0.400pt}}
\put(1429.0,275.0){\rule[-0.200pt]{2.409pt}{0.400pt}}
\put(281.0,322.0){\rule[-0.200pt]{2.409pt}{0.400pt}}
\put(1429.0,322.0){\rule[-0.200pt]{2.409pt}{0.400pt}}
\put(281.0,346.0){\rule[-0.200pt]{2.409pt}{0.400pt}}
\put(1429.0,346.0){\rule[-0.200pt]{2.409pt}{0.400pt}}
\put(281.0,357.0){\rule[-0.200pt]{4.818pt}{0.400pt}}
\put(261,357){\makebox(0,0)[r]{$10^{-4}$}}
\put(1419.0,357.0){\rule[-0.200pt]{4.818pt}{0.400pt}}
\put(281.0,392.0){\rule[-0.200pt]{2.409pt}{0.400pt}}
\put(1429.0,392.0){\rule[-0.200pt]{2.409pt}{0.400pt}}
\put(281.0,439.0){\rule[-0.200pt]{2.409pt}{0.400pt}}
\put(1429.0,439.0){\rule[-0.200pt]{2.409pt}{0.400pt}}
\put(281.0,463.0){\rule[-0.200pt]{2.409pt}{0.400pt}}
\put(1429.0,463.0){\rule[-0.200pt]{2.409pt}{0.400pt}}
\put(281.0,474.0){\rule[-0.200pt]{4.818pt}{0.400pt}}
\put(261,474){\makebox(0,0)[r]{$10^{-3}$}}
\put(1419.0,474.0){\rule[-0.200pt]{4.818pt}{0.400pt}}
\put(281.0,509.0){\rule[-0.200pt]{2.409pt}{0.400pt}}
\put(1429.0,509.0){\rule[-0.200pt]{2.409pt}{0.400pt}}
\put(281.0,556.0){\rule[-0.200pt]{2.409pt}{0.400pt}}
\put(1429.0,556.0){\rule[-0.200pt]{2.409pt}{0.400pt}}
\put(281.0,580.0){\rule[-0.200pt]{2.409pt}{0.400pt}}
\put(1429.0,580.0){\rule[-0.200pt]{2.409pt}{0.400pt}}
\put(281.0,591.0){\rule[-0.200pt]{4.818pt}{0.400pt}}
\put(261,591){\makebox(0,0)[r]{$10^{-2}$}}
\put(1419.0,591.0){\rule[-0.200pt]{4.818pt}{0.400pt}}
\put(281.0,626.0){\rule[-0.200pt]{2.409pt}{0.400pt}}
\put(1429.0,626.0){\rule[-0.200pt]{2.409pt}{0.400pt}}
\put(281.0,673.0){\rule[-0.200pt]{2.409pt}{0.400pt}}
\put(1429.0,673.0){\rule[-0.200pt]{2.409pt}{0.400pt}}
\put(281.0,696.0){\rule[-0.200pt]{2.409pt}{0.400pt}}
\put(1429.0,696.0){\rule[-0.200pt]{2.409pt}{0.400pt}}
\put(281.0,708.0){\rule[-0.200pt]{4.818pt}{0.400pt}}
\put(261,708){\makebox(0,0)[r]{$10^{-1}$}}
\put(1419.0,708.0){\rule[-0.200pt]{4.818pt}{0.400pt}}
\put(281.0,743.0){\rule[-0.200pt]{2.409pt}{0.400pt}}
\put(1429.0,743.0){\rule[-0.200pt]{2.409pt}{0.400pt}}
\put(281.0,790.0){\rule[-0.200pt]{2.409pt}{0.400pt}}
\put(1429.0,790.0){\rule[-0.200pt]{2.409pt}{0.400pt}}
\put(281.0,813.0){\rule[-0.200pt]{2.409pt}{0.400pt}}
\put(1429.0,813.0){\rule[-0.200pt]{2.409pt}{0.400pt}}
\put(281.0,825.0){\rule[-0.200pt]{4.818pt}{0.400pt}}
\put(261,825){\makebox(0,0)[r]{$10^{0~}\,$}}
\put(1419.0,825.0){\rule[-0.200pt]{4.818pt}{0.400pt}}
\put(281.0,860.0){\rule[-0.200pt]{2.409pt}{0.400pt}}
\put(1429.0,860.0){\rule[-0.200pt]{2.409pt}{0.400pt}}
\put(1320.0,123.0){\rule[-0.200pt]{0.400pt}{4.818pt}}
\put(1320,82){\makebox(0,0){ 0.018}}
\put(1320.0,840.0){\rule[-0.200pt]{0.400pt}{4.818pt}}
\put(1081.0,123.0){\rule[-0.200pt]{0.400pt}{4.818pt}}
\put(1081,82){\makebox(0,0){ 0.014}}
\put(1081.0,840.0){\rule[-0.200pt]{0.400pt}{4.818pt}}
\put(842.0,123.0){\rule[-0.200pt]{0.400pt}{4.818pt}}
\put(842,82){\makebox(0,0){ 0.01}}
\put(842.0,840.0){\rule[-0.200pt]{0.400pt}{4.818pt}}
\put(603.0,123.0){\rule[-0.200pt]{0.400pt}{4.818pt}}
\put(603,82){\makebox(0,0){ 0.006}}
\put(603.0,840.0){\rule[-0.200pt]{0.400pt}{4.818pt}}
\put(365.0,123.0){\rule[-0.200pt]{0.400pt}{4.818pt}}
\put(365,82){\makebox(0,0){ 0.002}}
\put(365.0,840.0){\rule[-0.200pt]{0.400pt}{4.818pt}}
\put(281.0,123.0){\rule[-0.200pt]{278.962pt}{0.400pt}}
\put(1439.0,123.0){\rule[-0.200pt]{0.400pt}{177.543pt}}
\put(281.0,860.0){\rule[-0.200pt]{278.962pt}{0.400pt}}
\put(281.0,123.0){\rule[-0.200pt]{0.400pt}{177.543pt}}
\put(100,491){\makebox(0,0){\rotatebox{90}{Relative amplitude $a(t)/a(0)$}}}
\put(860,21){\makebox(0,0){$1/\textrm{Re}\propto \nu$}}
\put(822,357){\makebox(0,0)[r]{at $t/T=5\ $}}
\put(842.0,357.0){\rule[-0.200pt]{24.090pt}{0.400pt}}
\put(1439,647){\usebox{\plotpoint}}
\multiput(1427.06,647.58)(-3.482,0.499){169}{\rule{2.877pt}{0.120pt}}
\multiput(1433.03,646.17)(-591.029,86.000){2}{\rule{1.438pt}{0.400pt}}
\multiput(830.58,733.58)(-3.332,0.498){105}{\rule{2.752pt}{0.120pt}}
\multiput(836.29,732.17)(-352.288,54.000){2}{\rule{1.376pt}{0.400pt}}
\multiput(473.19,787.58)(-3.182,0.495){35}{\rule{2.605pt}{0.119pt}}
\multiput(478.59,786.17)(-113.593,19.000){2}{\rule{1.303pt}{0.400pt}}
\multiput(353.52,806.59)(-3.564,0.482){9}{\rule{2.767pt}{0.116pt}}
\multiput(359.26,805.17)(-34.258,6.000){2}{\rule{1.383pt}{0.400pt}}
\multiput(313.52,812.61)(-4.258,0.447){3}{\rule{2.767pt}{0.108pt}}
\multiput(319.26,811.17)(-14.258,3.000){2}{\rule{1.383pt}{0.400pt}}
\multiput(293.52,815.61)(-4.258,0.447){3}{\rule{2.767pt}{0.108pt}}
\multiput(299.26,814.17)(-14.258,3.000){2}{\rule{1.383pt}{0.400pt}}
\put(1439,647){\circle{18}}
\put(842,733){\circle{18}}
\put(484,787){\circle{18}}
\put(365,806){\circle{18}}
\put(325,812){\circle{18}}
\put(305,815){\circle{18}}
\put(285,818){\circle{18}}
\put(892,357){\circle{18}}
\put(822,296){\makebox(0,0)[r]{at $t/T=10$}}
\put(842.0,296.0){\rule[-0.200pt]{24.090pt}{0.400pt}}
\put(1439,469){\usebox{\plotpoint}}
\multiput(1432.82,469.58)(-1.738,0.500){341}{\rule{1.488pt}{0.120pt}}
\multiput(1435.91,468.17)(-593.911,172.000){2}{\rule{0.744pt}{0.400pt}}
\multiput(836.08,641.58)(-1.661,0.499){213}{\rule{1.426pt}{0.120pt}}
\multiput(839.04,640.17)(-355.040,108.000){2}{\rule{0.713pt}{0.400pt}}
\multiput(478.39,749.58)(-1.575,0.498){73}{\rule{1.353pt}{0.120pt}}
\multiput(481.19,748.17)(-116.193,38.000){2}{\rule{0.676pt}{0.400pt}}
\multiput(359.05,787.58)(-1.703,0.492){21}{\rule{1.433pt}{0.119pt}}
\multiput(362.03,786.17)(-37.025,12.000){2}{\rule{0.717pt}{0.400pt}}
\multiput(319.84,799.59)(-1.484,0.485){11}{\rule{1.243pt}{0.117pt}}
\multiput(322.42,798.17)(-17.420,7.000){2}{\rule{0.621pt}{0.400pt}}
\multiput(299.05,806.59)(-1.756,0.482){9}{\rule{1.433pt}{0.116pt}}
\multiput(302.03,805.17)(-17.025,6.000){2}{\rule{0.717pt}{0.400pt}}
\put(1439,469){\circle*{18}}
\put(842,641){\circle*{18}}
\put(484,749){\circle*{18}}
\put(365,787){\circle*{18}}
\put(325,799){\circle*{18}}
\put(305,806){\circle*{18}}
\put(285,812){\circle*{18}}
\put(892,296){\circle*{18}}
\put(822,235){\makebox(0,0)[r]{at $t/T=15$}}
\put(842.0,235.0){\rule[-0.200pt]{24.090pt}{0.400pt}}
\put(1439,291){\usebox{\plotpoint}}
\multiput(1434.74,291.58)(-1.158,0.500){513}{\rule{1.026pt}{0.120pt}}
\multiput(1436.87,290.17)(-594.871,258.000){2}{\rule{0.513pt}{0.400pt}}
\multiput(837.92,549.58)(-1.106,0.500){321}{\rule{0.984pt}{0.120pt}}
\multiput(839.96,548.17)(-355.958,162.000){2}{\rule{0.492pt}{0.400pt}}
\multiput(480.06,711.58)(-1.065,0.499){109}{\rule{0.950pt}{0.120pt}}
\multiput(482.03,710.17)(-117.028,56.000){2}{\rule{0.475pt}{0.400pt}}
\multiput(361.09,767.58)(-1.061,0.495){35}{\rule{0.942pt}{0.119pt}}
\multiput(363.04,766.17)(-38.045,19.000){2}{\rule{0.471pt}{0.400pt}}
\multiput(321.26,786.58)(-1.017,0.491){17}{\rule{0.900pt}{0.118pt}}
\multiput(323.13,785.17)(-18.132,10.000){2}{\rule{0.450pt}{0.400pt}}
\multiput(301.26,796.58)(-1.017,0.491){17}{\rule{0.900pt}{0.118pt}}
\multiput(303.13,795.17)(-18.132,10.000){2}{\rule{0.450pt}{0.400pt}}
\put(1439,291){\makebox(0,0){$\times$}}
\put(842,549){\makebox(0,0){$\times$}}
\put(484,711){\makebox(0,0){$\times$}}
\put(365,767){\makebox(0,0){$\times$}}
\put(325,786){\makebox(0,0){$\times$}}
\put(305,796){\makebox(0,0){$\times$}}
\put(285,806){\makebox(0,0){$\times$}}
\put(892,235){\makebox(0,0){$\times$}}
\put(281.0,123.0){\rule[-0.200pt]{278.962pt}{0.400pt}}
\put(1439.0,123.0){\rule[-0.200pt]{0.400pt}{177.543pt}}
\put(281.0,860.0){\rule[-0.200pt]{278.962pt}{0.400pt}}
\put(281.0,123.0){\rule[-0.200pt]{0.400pt}{177.543pt}}
\end{picture}
  \caption{Relative amplitude $a(t)/a(0)$ against the inverse of the Reynolds number}\label{fig:wave_damping2}
\end{figure}
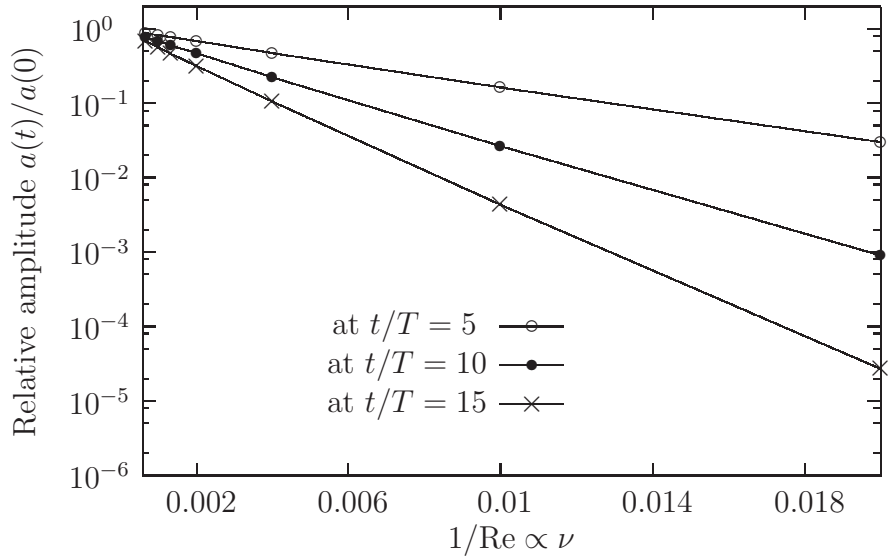

Using an energetic argument, Lamb \cite{lamb32:_hydrod} derived the approximate damping of a free wave due to viscosity as a function of time
\begin{equation}
  \label{eq:viscous_damping}
  a(t) = a(0) \textrm{e}^{-2\nu k^2 t}.
\end{equation}
Figure \ref{fig:wave_damping} displays the computed relative wave amplitude $a(t)/a(0)$ with respect to the simulation time $t/T$. The excellent linear fits obtained for the seven values of the Reynolds number are in perfect agreement with the exponential viscous damping. Equation \eqref{eq:viscous_damping} also shows that the relative wave amplitude is proportional to the kinematic viscosity, if plotted in $y$-log scale. This second point is verified in Figure \ref{fig:wave_damping2}, where $a(t)/a(0)$ is plotted against the inverse of the Reynolds number which is by definition proportional to the kinematic viscosity $\nu$.

Those results are evidences of the robustness and accuracy of our moving-grid technique in handling large-deformations for moving-boundary problems such as the one considered here.

\section{Conclusions}
\label{sec:conclusions}
A numerical model for solving two- and three-dimensional moving-boundary problems such as free-surface flows or fluid-structure interaction is proposed. This model relies on a moving-grid technique to solve the Navier--Stokes equations expressed in the arbitrary Lagrangian-Eulerian kinematics and discretized by the spectral element method. A detailed analysis of the continuous and discretized formulations of the general problem in the ALE frame, with non-homogeneous and unsteady boundary conditions is presented. Particular emphasis was put on the weak formulation and its semi-discrete counterpart. The moving-grid algorithm which is one of the key ingredient of our numerical model, is based on the computation of the ALE mesh velocity with the same accuracy and numerical technique as the fluid velocity. The coupling between the Navier--Stokes computation and the one for the mesh velocity is effective through the problem boundary conditions. It is noteworthy that the coupling in the interior Navier--Stokes computation is effective through the modified convective term which is induced by what is happening at the boundaries. Three numerical test results are presented in the two particular cases of interest, namely fluid-structure interactions and free-surface flows. First the influence of the deformation of the grid on the accuracy of the numerical model is evaluated. In a second problem, two motions (translation and rotation) of a cylinder immersed in a fluid is computed. Lastly, large-amplitude sloshing in a three-dimensional tank is simulated. The results obtained are showing very good with the theoretical results when available, therefore leading to a validation of our numerical model.
\ack
This research is being partially funded by a Swiss National Science Foundation Grant (No. 200020--101707), whose support is gratefully acknowledged.

\end{document}